\documentclass[american,  aps,pra,reprint,superscriptaddress, longbibliography]{revtex4-1}
\usepackage[T1]{fontenc}
\usepackage[latin9]{inputenc}
\usepackage{color}
\usepackage{babel}
\usepackage{amsthm}
\usepackage{amsmath}
\usepackage{amssymb}
\usepackage{graphicx}
\usepackage[unicode=true,pdfusetitle, bookmarks=true,bookmarksnumbered=false,bookmarksopen=false,  breaklinks=false,pdfborder={0 0 0},backref=false,colorlinks=false] {hyperref}
\hypersetup{ colorlinks,linkcolor=myurlcolor,citecolor=myurlcolor,urlcolor=myurlcolor}

\makeatletter
\@ifundefined{textcolor}{}
{%
 \definecolor{BLACK}{gray}{0}
 \definecolor{WHITE}{gray}{1}
 \definecolor{RED}{rgb}{1,0,0}
 \definecolor{GREEN}{rgb}{0,1,0}
 \definecolor{BLUE}{rgb}{0,0,1}
 \definecolor{CYAN}{cmyk}{1,0,0,0}
 \definecolor{MAGENTA}{cmyk}{0,1,0,0}
 \definecolor{YELLOW}{cmyk}{0,0,1,0}
 }
\theoremstyle{plain}

\theoremstyle{plain}

\ifx\proof\undefined
\newenvironment{proof}[1][\protect\proofname]{\par
\normalfont\topsep6\p@\@plus6\p@\relax
\trivlist
\itemindent\parindent
\item[\hskip\labelsep
\scshape
#1]\ignorespaces
}{%
\endtrivlist\@endpefalse
}
\providecommand{\proofname}{Proof}
\fi
\theoremstyle{plain}

\providecommand{\lemmaname}{Lemma}
\providecommand{\definitionname}{Definition}
\providecommand{\propositionname}{Proposition}

\usepackage{babel}
\usepackage{txfonts}%\usepackage{braket}
\usepackage{colortbl}\definecolor{myurlcolor}{rgb}{0,0,0.7}
\newcommand{\tr}{{\operatorname{Tr\,}}}

\def\ket#1{| #1 \rangle}
\def\bra#1{\langle  #1 |}

\def\proj#1{| #1 \rangle\!\langle #1 |}

\DeclareMathOperator{\Tr}{\mathrm{Tr}}%trace2
\newcommand{\haH}

\definecolor{orange}{RGB}{255,127,0}

\begin{document}

\title{Steady-state Quantum Thermodynamics with Synthetic Negative Temperatures  }

\author{Mohit Lal Bera}
\affiliation{ICFO - Institut de Ci\`encies Fot\`oniques, The Barcelona Institute of Science and Technology, 08860 Castelldefels (Barcelona), Spain.}

\author{Tanmoy Pandit}
\email{tanmoypandit163@gmail.com}
\affiliation{Fritz Haber Research Center for Molecular Dynamics, Hebrew University of Jerusalem, Jerusalem 9190401, Israel}

\author{Kaustav Chatterjee}
\affiliation{Faculty of Physics, Arnold Sommerfeld Centre for Theoretical Physics (ASC),
Ludwig-Maximilians-Universit\"{a}t M\"{u}nchen, Theresienstr. 37, 80333 M\"{u}nchen, Germany}

\author{Varinder Singh}
\affiliation{Center for Theoretical Physics of Complex Systems,
Institute for Basic Science (IBS), Daejeon 34126, Korea}

\author{Maciej Lewenstein}
\affiliation{ICFO - Institut de Ci\`encies Fot\`oniques, The Barcelona Institute of Science and Technology, 08860 Castelldefels (Barcelona), Spain.}
\affiliation{ICREA, Pg. Lluis Companys 23, ES-08010 Barcelona, Spain.}

\author{Utso Bhattacharya}
\affiliation{ICFO - Institut de Ci\`encies Fot\`oniques, The Barcelona Institute of Science and Technology, 08860 Castelldefels (Barcelona), Spain.}

\author{Manabendra Nath Bera}
\email{mnbera@gmail.com}
\affiliation{Department of Physical Sciences, Indian Institute of Science Education and Research (IISER), Mohali, Punjab 140306, India}

\begin{abstract} 
A bath with a negative temperature is a subject of intense debate in recent times. It raises fundamental questions not only on our understanding of negative temperature of a bath in connection with thermodynamics but also on the possibilities of constructing devices using such baths. In this work, we study steady-state quantum thermodynamics involving baths with negative temperatures. A bath with a negative temperature is created synthetically using two baths of positive temperatures and weakly coupling these with a qutrit system. These baths are then coupled to each other via a working system. At steady-state, the laws of thermodynamics are analyzed. We find that whenever the temperatures of these synthetic baths are identical, there is no heat flow, which reaffirms the zeroth law. There is always a spontaneous heat flow for different temperatures. In particular, heat flows from a bath with a negative temperature to a bath with a positive temperature which, in turn, implies that a bath with a negative temperature is `hotter' than a bath with a positive temperature. This warrants an amendment in the Kelvin-Planck statement of the second law, as suggested in earlier studies. In all these processes, the overall entropy production is positive, as required by the Clausius statement of the second law. We construct continuous heat engines operating between positive and negative temperature baths. These engines yield maximum possible heat-to-work conversion efficiency, that is, unity. We also study the thermodynamic nature of heat from a bath with a negative temperature and find that it is thermodynamic work but with negative entropy. 
\end{abstract}

\maketitle
\normalfont

\section{Introduction}
Thermodynamics constitutes a fundamental building block of the modern understanding of nature. With the advent of quantum mechanics, there have been numerous efforts to extend the framework to systems composed of a finite or large number of quantum particles while each particle has a discrete energy spectrum and the states are in a superposition of different energy levels, see for example \cite{Binder18}. One of the possibilities for quantum systems with bounded energy is that, in certain situations, they can assume `negative' temperatures. This arises when the population distribution of a system or bath becomes an inverted Boltzmann distribution, i.e., states with higher energy are populated more than the ones with lower energy. It was first pointed out by Purcell and Pound in the context of nuclear spin systems \cite{Purcell1951}. Subsequently, Ramsey comprehensively discussed the thermodynamic implications of such negative temperatures and the inter-relation between negative and positive temperatures \cite{Ramsey1956}. He advocates for an amendment to the Kelvin-Planck statement of the second law to incorporate that heat flows spontaneously from a bath with a negative temperature to one with a positive temperature. In this sense, the negative temperature is `hotter' than a positive temperature.

Initially, Sch\"{o}pf raised some foundational questions regarding the dynamics of negative temperature \cite{Schpf1962}. He claimed that it is impossible to transform a thermodynamic system adiabatically: from a positive finite temperature to the positive infinite temperature, then from there to a negative infinite (Boltzmann) temperature, and then subsequently to a negative finite (Boltzmann) temperature \cite{Schpf1962}. Tykodi and Tremblay \cite{Tykodi1975, Tremblay1976, Tykodi1978} disagreed and showed that the arguments used by Sch\"{o}pf are thermodynamically inconsistent as these violate the second law of thermodynamics.
Nevertheless, the debates on thermodynamics with negative temperatures are not settled. Recent theoretical \cite{Mosk2005, Rapp2010} and experimental \cite{Braun2013, Carr2013, Muniz23} studies with cold atoms have brought the debate on negative temperatures back again into the spotlight. The works in \cite{Dunkel2013, Hilbert2014, Hanggi2016} claim that ``all previous negative temperature claims and their implications are invalid as they arise from the use of an entropy definition that is inconsistent both mathematically and thermodynamically." Another study in \cite{Struchtrup2018} states that thermodynamic equilibrium at negative temperatures would be unstable but can be used for work storage or battery. Several researchers have come forward and systematically explained that identification of the thermodynamic entropy exclusively with the volume entropy proposed by Gibbs is the root of all doubts \cite{Cerino2015, Buonsante2016, Poulter2016, Abraham2017, Swendsen2018}, and it is inconsistent with the postulates of thermodynamics \cite{Wang2015, Swendsen2016}. Using Boltzmann entropy as the thermodynamic entropy, they argue that negative temperature is a valid extension of thermodynamics.

Apart from these foundational issues, there are questions on whether a negative temperature bath can be used to construct thermal devices, such as heat engines, refrigerators, heat pumps, etc. Initially, a study on Carnot engines was made by Geusic, et al. \cite{Geusic1967} and, later on, by Landsberg and Nakagomi \cite{Landsberg1977, Nakagomi1980} in this context. There are also some studies about how the Carnot cycle should be modified in the presence of negative temperatures \cite{DunningDavies1976, DunningDavies1978,Landsberg1980}. Further, some propositions are made to construct a quantum Otto engine \cite{Xi2017, deAssis2019} and refrigerators \cite{Damas2022} using a bath with effective negative temperature. It is shown that the heat-to-work conversion efficiency of an engine operating between negative and positive temperatures would be greater than unity \cite{Braun2013, Carr2013, deAssis2019}. However, much of these models of thermal devices either utilize already existing negative temperature baths without caring how it may be created or effectively prepare one by inverting
the populations using some external means.

In this article, we outline how to create a thermal bath with arbitrary temperature, including negative temperature, and study steady-state quantum thermodynamics. The bath is synthesized by letting a quantum system interact simultaneously with two thermal baths at different positive temperatures without external driving. We study various laws of steady-state quantum thermodynamics with these synthetic baths and construct continuous heat engines. We start by proving the zeroth law and show that there is no net heat flow whenever two such baths with identical temperatures are brought in contact with each other. This, in turn, legitimizes the notion of the temperature of a synthetic bath (namely, the synthetic temperature). In the case of two different temperatures, we prove the Kelvin-Planck statement of the second law and demonstrate that there is a spontaneous heat flow from a bath with a negative temperature to one with a positive temperature. This corroborates with the finding of Ramsey \cite{Ramsey1956} - baths with negative temperatures are `hotter' than the ones with positive temperatures. Interestingly, in such cases, the entropy flow is opposite to the direction of heat flow which is again expected for the baths with negative temperatures. We also construct Carnot engines involving synthetic baths and find that engines operating between positive and negative temperature baths can yield unit engine efficiency. This leads us to question the physical meaning of heat flow in the presence of a bath with negative temperatures. With a systematic analysis, we show that the heat associated with a bath with a negative temperature is equivalent to work but with a negative entropy flow. 

\section{Synthetic baths and Negative temperatures \label{Sec:SynthBath}}
In general, naturally occurring thermal equilibrium results in non-negative temperatures. Only in certain situations, as discussed earlier, can the temperatures be negative. Below, we introduce a method through which a bath can be synthesized. The temperatures of these synthetic baths can assume arbitrary values, including negative ones.

\begin{figure}
\includegraphics[width=0.85\columnwidth]{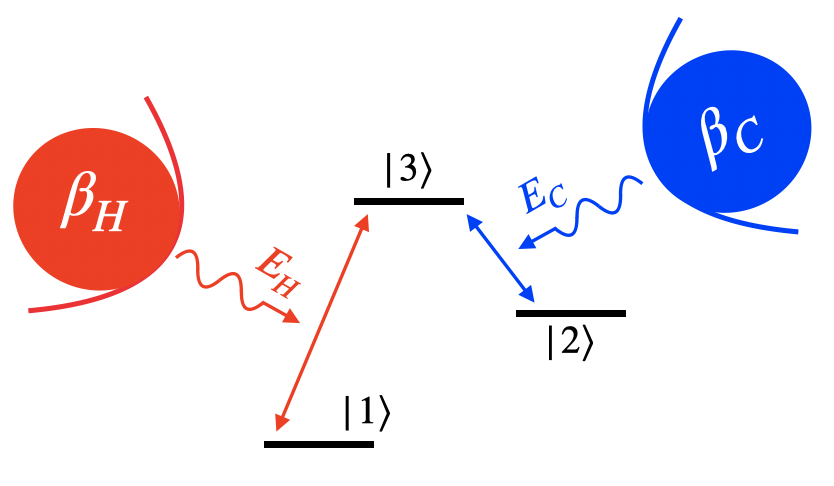}
\caption{A synthetic bath is created using two baths with different temperatures and letting them interact with a qutrit system. In particular, the hot bath ($H$) with inverse temperature $\beta_H$ is weakly coupled to the energy eigenstates $\ket{1}$ and $\ket{3}$. The cold bath ($C$) with inverse temperature $\beta_C$ weakly interacts with the energy eigenstates $\ket{2}$ and $\ket{3}$. As a result, the populations of the states $\ket{1}$ and $\ket{2}$ reach an equilibrium corresponding to a synthetic temperature $\beta_S$. By tuning the temperature of the baths and the energy spacing between the states, an arbitrary synthetic temperature can be obtained, including negative temperatures. See text for more details.} \label{fig:SynthBath}
\end{figure}

The method utilizes a qutrit system, a hot bath ($H$) with inverse temperature $\beta_H$, and a cold bath ($C$) at inverse temperature $\beta_C$. The energy levels of the qutrit are denoted by $\ket{1}$, $\ket{2}$, $\ket{3}$, with the corresponding Hamiltonian $H=(E_H - E_C) \ \proj{2}+ E_H \ \proj{3}$. As shown in Fig.~\ref{fig:SynthBath}, the hot (cold) bath weakly interacts with the levels $\ket{1}$ and $\ket{3}$ (levels $\ket{2}$ and $\ket{3}$). As convention, we consider $\beta_H < \beta_C$, Planck constant $\hbar=1$, and the Boltzmann constant $k_B=1$ throughout the article. The levels $\ket{1}$ and $\ket{2}$ are not directly coupled. However, they are indirectly linked through the level $\ket{3}$. When the couplings between the qutrit and the baths are weak and satisfy the Markov condition, the overall dynamics is expressed in terms of the Lindblad or Lindblad-Gorini-Kossakowski-Sudarshan (LGKS) form of master equation \cite{Breuer2007}
\begin{align}
   \dot{\rho}= \mathcal{L}_U (\rho) + \mathcal{L}_H (\rho)  + \mathcal{L}_C (\rho). \label{eq:MasterEqnQutrit}
\end{align}
Here $\rho$ represents the density matrix corresponding to a state of the qutrit. The first term on the right-hand side of Eq.~\eqref{eq:MasterEqnQutrit} $\mathcal{L}_U (\rho)=i \ [\rho, H]$ takes care of the unitary part of the evolution due to the system Hamiltonian $H$. The second and third terms, with the Lindblad super-operators (LSOs) $\mathcal{L}_H(\cdot)$ and $\mathcal{L}_C(\cdot)$, represent the contributions due to the dissipative part of the evolution induced by the hot and cold thermal baths, respectively. The LSOs are expressed (for $X=H, \ C$) as
\begin{align}
   \mathcal{L}_X(\rho)= \  &\Gamma_X (N_X +1) \left( A_X \rho A_X^\dagger  -1/2 \ \{A_X^\dagger A_X, \rho \} \right) \nonumber \\ &+ \Gamma_X N_X   \left( A_X^\dagger \rho A_X -1/2 \ \{ A_X A_X^\dagger, \rho \} \right),
\end{align}
where $A_H= |1\rangle \langle 3|$, $A_C= |2\rangle \langle 3|$, anti-commutator $\{Y,Z \}=YZ + ZY$, and $N_X=1/(e^{\beta_X E_X} -1)$. The coefficient $\Gamma_X$ is the Weiskopf-Wigner decay constant. The overall dynamics leads to heat exchange between the baths and the system. The heat fluxes are quantified as $\dot{Q}_X=  \Tr[\mathcal{L}_X(\rho) \ H]$ due to interaction with the bath $X$ \cite{Boukobza2006a, Boukobza2006b, Boukobza2007}. Heat flux $\dot{Q}_X >0$ implies that heat is flowing into the qutrit system from the bath with inverse temperature $\beta_X$.

This dynamics always leads to a steady state, say $\sigma$, which is diagonal in the energy eigenstates. The the populations $\{p_i\}$ of the states $\{\ket{i}\}$ satisfy
\begin{align*}
    \frac{p_1}{p_3}=e^{\beta_H E_H}, \ \ \mbox{and} \ \ \frac{p_2}{p_3}=e^{\beta_C E_C}.
\end{align*}
The populations corresponding to states $\ket{1}$ and $\ket{3}$ attain thermal equilibrium with the hot bath, and similarly, the populations of $\ket{2}$ and $\ket{3}$ attain thermal equilibrium with the cold bath. In fact, the dynamics drives the overall system to reach thermal equilibrium, albeit in interactions with two baths at different temperatures. This is justified because the heat flux and entropy production vanish, i.e., $\dot{Q}_X=0$ for $X= H, C$ and $-\beta_H \dot{Q}_H -\beta_C \dot{Q}_C=0$ respectively. As the entire system is in thermodynamic equilibrium, so are the populations of the states $\ket{1}$ and $\ket{2}$.

In general, if one introduces an interaction between the levels $\ket{1}$ and $\ket{2}$, be it time-dependent or time-independent, the heat and entropy fluxes become non-zero \cite{Boukobza2006b, Boukobza2007} and the corresponding populations change. But once the interaction is switched off, the populations revert to their equilibrium values. This is as if the levels $\ket{1}$ and $\ket{2}$, or the subspace spanned by these two levels, are interacting with a synthetic bath at inverse temperature $\beta_S$, defined as
\begin{align}
\beta_S= \frac{1}{E_S} \ln\left(\frac{p_1}{p_2}\right) = \frac{1}{E_S} \ln\left(\frac{p_1}{p_3} \ \frac{p_3}{p_2} \right) = \frac{\beta_H E_H - \beta_C E_C}{E_S}, \label{eq:SynthTemp}
\end{align}
where $E_S=E_H-E_C$. We note that a Lindblad super-operator (LSO) cannot be given exclusively for the equilibration dynamics due to the synthetic bath. However, as we discuss in the later sections, this is a legitimate thermal bath. We call $\beta_S$ the 'synthetic' inverse temperature because it can be tuned to assume arbitrary values, including negative values, by changing the energy-level spacings and the $\beta_H$ and $\beta_C$. In literature, there are debates on whether the temperature of a system can be continuously changed from a positive to a negative equilibrium temperature \cite{Struchtrup2018}. However, in this setup, the inverse temperature of the synthetic bath can be tuned continuously, e.g., from $\beta_S>0$ to $\beta_S <0$, including $\beta_S=0$ (infinite temperature).

\section{Thermodynamics with synthetic baths}
\begin{figure} 
		\includegraphics[width=1\columnwidth]{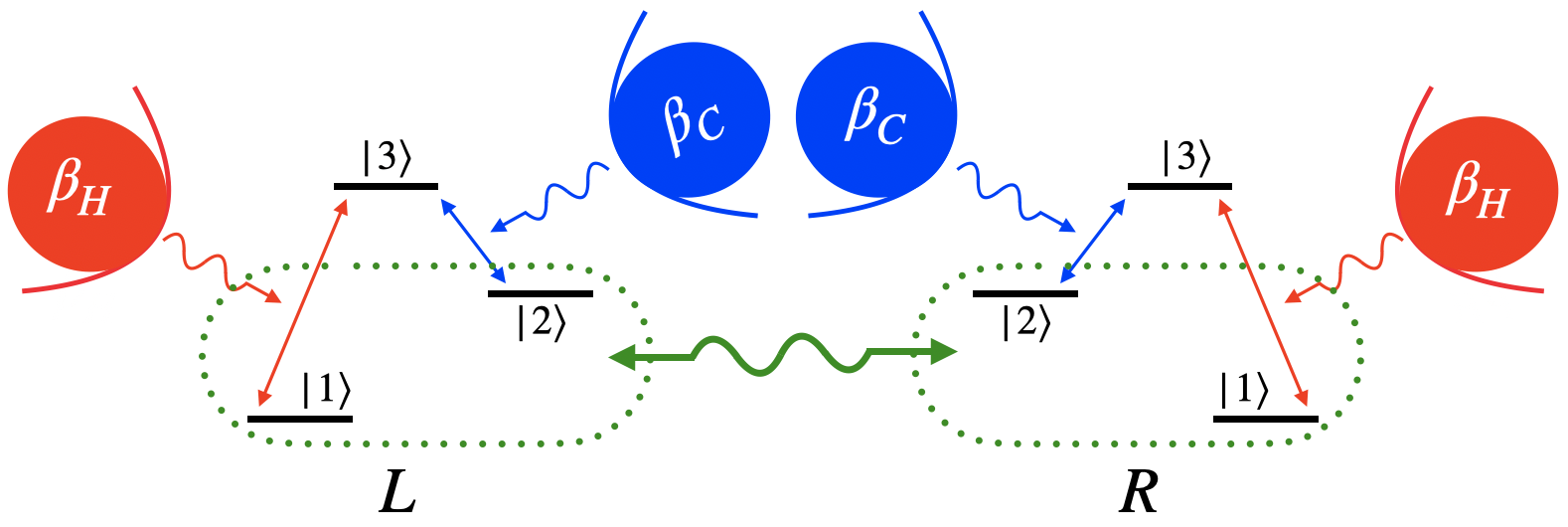}
\caption{\label{fig:2SynthBaths} Two synthetic baths are engineered with the help of a hot and a cold bath at inverse temperatures $\beta_H$ and $\beta_C$, respectively. Each synthetic bath is created by letting the baths weakly interact with one qutrit, as shown in Fig.~\ref{fig:SynthBath}. Different synthetic temperatures are engineered by tuning energy spacings between the states $\ket{1}$, $\ket{2}$, and $\ket{3}$. In addition, an interaction is introduced between the synthetic baths to study the heat and entropy flow. See text for more details.  }
	\end{figure}

To study thermodynamics with synthetic temperatures, we consider two different qutrit systems $L$ and $R$ with the corresponding Hamiltonians $H_X=(E_{XH} - E_{XC}) \ \proj{2}+E_{XH} \ \proj{3}$, with $X=L, R$. We assume $E_{XH} - E_{XC}=E_S$ for both systems, i.e., the energy spacing between $\ket{1}$ and $\ket{2}$ for both $L$ and $R$ are same. For brevity, we denote $H_L \equiv H_L \otimes \mathbb{I}$ and $H_R \equiv \mathbb{I} \otimes H_R$. Each system couples a hot and a cold bath with inverse temperatures $\beta_H$ and $\beta_C$ respectively (see Fig.~\ref{fig:2SynthBaths}) and reaches an equilibrium state. Without an interaction in between, the equilibrium state of the composite $LR$ becomes
\begin{align}
    \rho_{L} \otimes \rho_R=\sum_{m, n=1}^3 p_m q_n \  \proj{m \ n}, \label{eq:IniLRstate}
\end{align}
where $p_1/p_3=e^{\beta_H E_{LH}}$, $p_2/p_3=e^{\beta_C E_{LC}}$, $q_1/q_3=e^{\beta_H E_{RH}}$, and $q_2/q_3=e^{\beta_C E_{RC}}$. The population ratio between the degenerate energy states $\ket{21}$ and $\ket{12}$ is
\begin{align}
 p_1q_2/p_2q_1=e^{(\beta_{LS}-\beta_{RS})E_S}, \label{eq:IniPopRatio}
 \end{align}
where $\beta_{LS}$ and $\beta_{RS}$ are the synthetic temperatures corresponding to qutrit $L$ and $R$ respectively.

An interaction is introduced that only couples subspace spanned by the energy levels belonging to $\ket{1}$ and $\ket{2}$ in each qutrit, ensuring an energy exchange between $L$ and $R$ only through these subspaces. The most general interaction Hamiltonian that drives an energy exchange between these subspaces is given by
\begin{align}
H_{in}= (\lambda + i \ \gamma) \ \ket{12}\bra{21} + (\lambda - i \ \gamma) \ \ket{21}\bra{12}, \label{eq:IntHamTwoQutrit}
\end{align}
where $\lambda, \gamma \in \mathbb{R}$. This interaction also strictly conserves energy, as $[H_{in}, H_L + H_R]=0$. The overall dynamics of $LR$ is expressed as
\begin{align}
\dot{\rho}_{LR}= i \  [\rho_{LR}, H_T] + \mathcal{L}_L (\rho_{LR}) + \mathcal{L}_R (\rho_{LR}), \label{eq:2QutritME}
\end{align}
where $H_T=H_L+H_R+H_{in}$, $\mathcal{L}_L(\cdot)=\mathcal{L}_{LH}(\cdot)+\mathcal{L}_{LC}(\cdot)$, and $\mathcal{L}_R(\cdot)=\mathcal{L}_{RH}(\cdot) + \mathcal{L}_{RC}(\cdot)$. Here, $\mathcal{L}_{XH}(\cdot)$ and $\mathcal{L}_{XC}(\cdot)$ are the LSOs taking into account the dissipative part of the dynamics due to the coupling with hot and cold baths, respectively, with the qutrit $X$. Under this dynamics, the composite system $LR$ reaches a steady state, say $\sigma_{LR}$. Then, the heat flux and the entropy flux, respectively, are
\begin{align}
    \dot{Q}_X=  \Tr[\mathcal{L}_X(\sigma_{LR}) \ H_{X}], \ \ \
     \dot{S}_X= \beta_{XS} \dot{Q}_X,
\end{align}
with $X=L, R$, and $\dot{Q}_X=\dot{Q}_{XH} + \dot{Q}_{XC}$.
%, and at steady state $\dot{Q}_l = - \dot{Q}_r$.

In absence of any interaction between the $L$ and $R$, the steady (or equilibrium) state is $\sigma_{LR}=\rho_L \otimes \rho_R$ (see Eq.~\eqref{eq:IniLRstate}). Then, the heat flux from $L$ is $\dot{Q}_L=0$, as $\dot{Q}_{LH}= \dot{Q}_{LC}=0$. However in presence of interaction via $H_{in}$, the steady state becomes $\sigma_{LR} \neq \rho_L \otimes \rho_R$, and then the $\dot{Q}_L = \dot{Q}_{LH} + \dot{Q}_{LC}\neq 0$. This means that there is heat flux through the subspace spanned by $\{\ket{1}, \ket{2}\}$ of $L$, which we may consider as the heat flux due to the synthetic bath associated with $L$. By convention, $\dot{Q}_L > 0$ implies a heat flux from the synthetic bath to $L$, which is then passed to $R$. With these tools at hand, we now set out to explore steady-state quantum thermodynamics with synthetic temperatures. Note that the first law is always respected at steady state as $\dot{Q}_L+ \dot{Q}_R=0$. Thus, our emphasis would be on studying the zeroth and second laws.

\ 

\begin{figure*}
		\includegraphics[width=0.97\columnwidth]{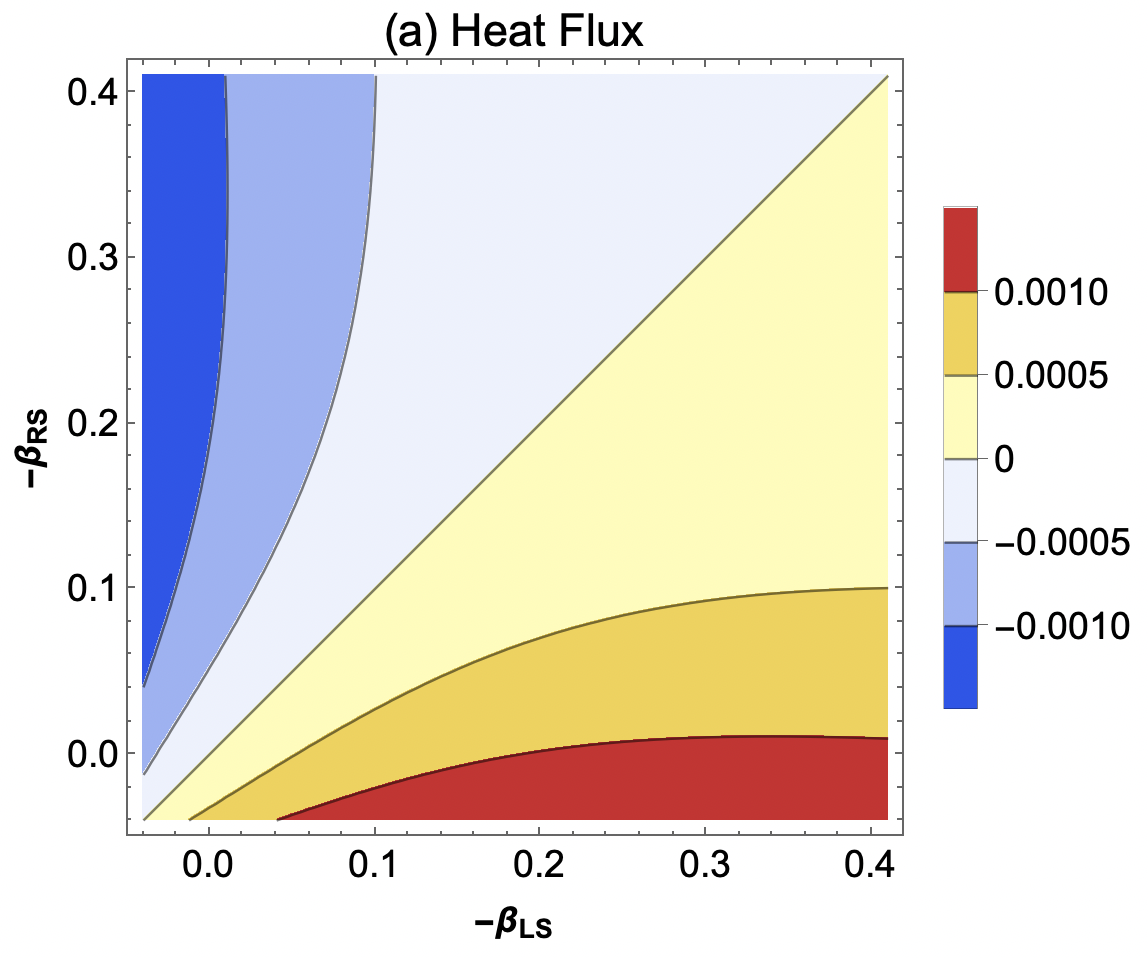}
  \includegraphics[width=1\columnwidth]{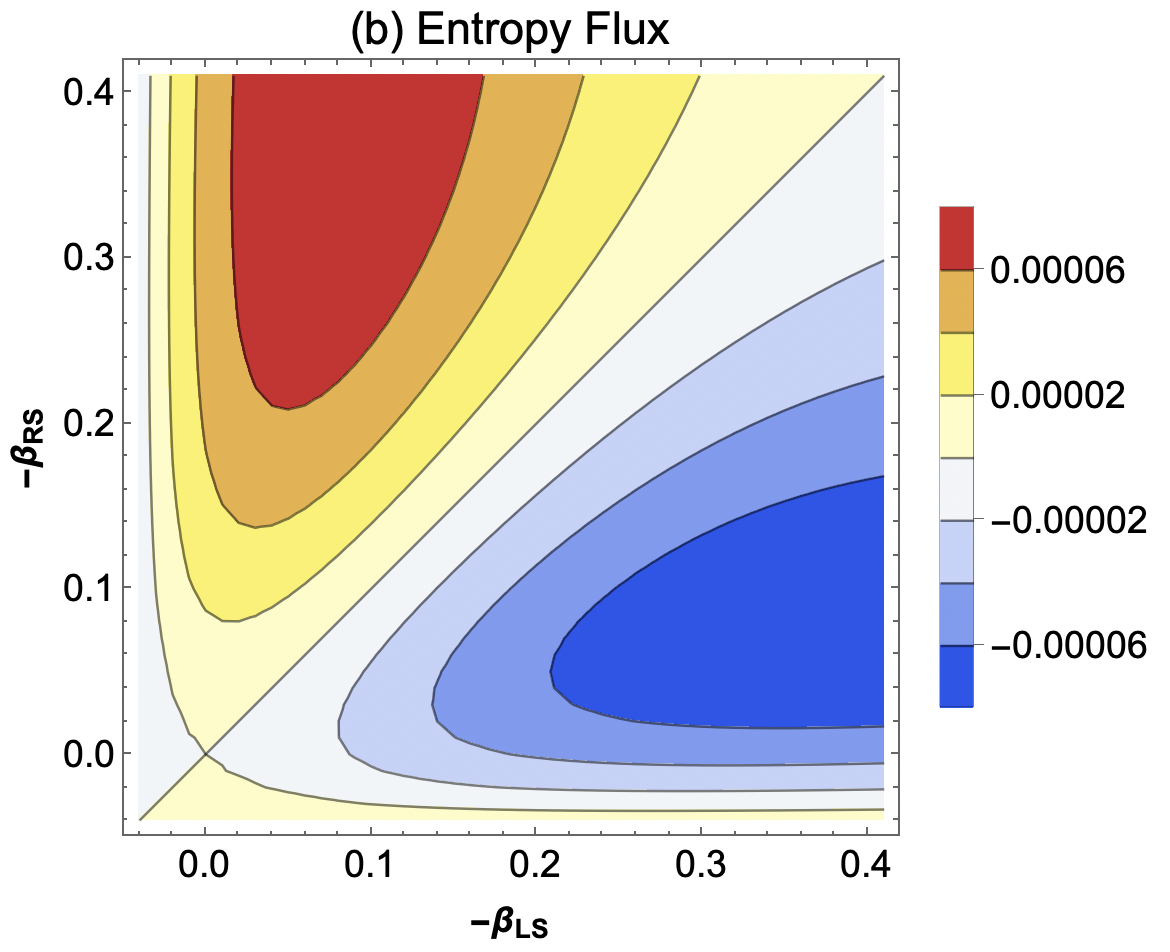}
\caption{The figures represent heat and entropy fluxes, $\dot{Q}_L$ and $\dot{S}_L$, respectively, through the qutrit $L$. The numerical calculation is carried out with the parameters:  $\Gamma_{H}=\Gamma_{C}=0.001$, $\beta_H=0.05$, $\beta_{C}=1$, $\lambda=1$, $\gamma=0$, $E_S=9.5$. The different synthetic inverse temperatures are obtained by tuning the energy $E_H$ of $\ket{3}$ for the qutrits $L$ and $R$. (a) The density plot represents the Kelvin-Planck statement of the second law in terms of the heat flux through $L$, i.e., $\dot{Q}_L$.  The plot shows no heat flux for $\beta_{LS}=\beta_{RS}$. This corroborates with the zeroth law. However, there is a positive (negative) heat flux $\dot{Q}_L$, i.e., heat flows from $L$ to $R$ (from $R$ to $L$), whenever $-\beta_{LS} > -\beta_{RS}$ ($-\beta_{LS} < -\beta_{RS}$) implying bath with $\beta_{LS}$ ($\beta_{RS}$) is hotter than $\beta_{RS}$ ($\beta_{LS}$). Clearly, a bath with a negative temperature is always `hotter' than any bath with a positive temperature. (b) The density plot represents the variation of entropy flux $\dot{S}_L$ for different synthetic temperatures. As seen from the plot, for $\beta_{LS}=\beta_{RS}$, $\dot{S}_L=0$. For $\beta_{LS}>0$ and $\beta_{RS}>0$, the direction of entropy flux is same with the heat flux as expected for the baths with positive temperatures. However, the direction of heat flow is opposite to the direction of entropy flow, in general, for baths with negative inverse temperatures. See text for details.} \label{fig:KP}
\end{figure*}

{\bf Zeroth law} - 
In thermodynamics, the zeroth law interlinks the notion of temperature with equilibrium. It states that if two systems are in thermal equilibrium, they must have the same temperature and vice versa. Again, thermal equilibrium implies that when two systems are in contact, there is no net flux in any thermodynamic quantities (such as heat and entropy) between systems. In such situations, the overall entropy production also vanishes. Below we show that, in the two-qutrit scenario discussed above, there is no net flux of any thermodynamic quantity whenever the synthetic temperatures are identical for both qutrits.

Recall the setup we consider in Fig.~\ref{fig:2SynthBaths}. Without any interaction, the steady state of $LR$ is $\rho_{L} \otimes \rho_R$, which is diagonal in the energy eigenstates (see Eq.~\eqref{eq:IniLRstate}). With same synthetic inverse temperatures $\beta_{LS}=\beta_{RS}$, the populations of the states $\ket{12}$ and $\ket{21}$ satisfy $p_1q_2=p_2q_1$ (see Eq.~\eqref{eq:IniPopRatio}). It means, the matrix corresponding to the state $\rho_{L} \otimes \rho_R$ in the subspace spanned by $\{ \ket{12}, \ \ket{21} \}$ is proportional to an identity operator. Now an interaction between $L$ and $R$ introduced by $H_{in}$, as in Eq.~\eqref{eq:IntHamTwoQutrit}. It is easily seen that $[H_{in}, \rho_{L} \otimes \rho_R]=0$ for $\beta_{LS}=\beta_{RS}$. Thus, even after the interaction is switched on, the steady state remains unaltered, i.e., $\sigma_{LS}=\rho_{L} \otimes \rho_R$ and $\dot{\sigma}_{LR}=\bold{0}$. Hence, there is no exchange of heat and entropy between $L$ and $R$, as $\dot{Q}_X=\dot{S}_X=0$ for $X=L, R$. This implies that the synthetic baths are in thermal equilibrium whenever the synthetic temperatures are identical, irrespective of whether the temperatures are positive or negative. A numerical analysis also confirms this. See Fig.~\ref{fig:KP}(a). \\

{\bf Second law} - 
For $\beta_{LS} \neq \beta_{RS}$, heat and entropy flow is possible from one qutrit to the other. However, the flow cannot be arbitrary. The second law dictates the physically allowed processes given that zeroth and first laws are respected. There are various statements of the second law. Below, we analyze the Kelvin-Planck statement to deal with the directionality of heat flow and the Clausius statement regarding entropy production.

Let us consider the case for which $-\beta_{LS}>-\beta_{RS}$. When there is no interaction between the qutrits, the equilibrium state of the composite $LS$ is $\rho_L \otimes \rho_R$. The state is expressed in the block-diagonal form as
\begin{align*}
\rho_L \otimes \rho_R=  \Pi_0 \rho_L \otimes \rho_R \Pi_0 + \Pi_1 \rho_L \otimes \rho_R \Pi_1,    
\end{align*}
where $\Pi_0=\proj{12} + \proj{21}$ and $\Pi_1=\mathbb{I}-\Pi_0$. The populations corresponding to the energy eigenstates $\ket{12}$ and $\ket{21}$ satisfy $p_1q_2 < p_2q_1$ (see Eq.~\eqref{eq:IniPopRatio}).

Now let us disconnect the thermal baths and introduce an interaction driven by $H_{in}$. This, or the total Hamiltonian $H_T$, evolves the composite and induces a rotation onto the subspace spanned by $\ket{12}$ and $\ket{21}$ only. At the same time, the other part of the density matrix remains unchanged. As a result, there appear off-diagonal elements in this subspace. Say, the state of $LR$ after any evolution becomes 
\begin{align*}
 \rho_{LR}^\prime =a \  \ket{1 2}\bra{1  2} & + b \  \ket{1 2}\bra{2 1} + c \  \ket{2 1}\bra{1 2} + d \  \ket{2 1}\bra{2 1} \\  &+ \Pi_1 \rho_L \otimes \rho_R \Pi_1.
\end{align*}
The unitary nature of the evolution in the subspace $\ket{12}$ and $\ket{21}$ guarantees that $a > p_1q_2$ and $d < p_2q_1$. For this reason and as the off-diagonal elements do not contribute to the populations of the reduced state of $L$, i.e., $\rho_L^\prime = \Tr_{R} \rho_{LR}^\prime$, we find $p^\prime_1 = \bra{1}\rho_L^\prime \ket{1} > p_1$ and $p^\prime_2 = \bra{2}\rho_L^\prime \ket{2} < p_2$. Similarly, for the reduced state of $\rho_R^\prime = \Tr_{L} \rho_{LR}^\prime$, the modified populations becomes $q^\prime_1 = \bra{1}\rho_R^\prime \ket{1} < q_1$ and $q^\prime_2 = \bra{2}\rho_R^\prime \ket{2} > q_2$. Note, the populations corresponding to level $\ket{3}$ for both $L$ and $R$ remain unchanged, i.e., $\bra{3}\rho_L^\prime \ket{3}= p_3$ and $\bra{3}\rho_R^\prime \ket{3}= q_3$. Clearly, the qutrit $L$ loses some energy. As the evolution respects strict energy conservation, the qutrit $R$ gains the same amount of energy. Thus, any evolution due to $H_{in}$ ensures that there is an energy flow from $L$ to $R$ for $-\beta_{LS}>-\beta_{RS}$. After this modification, if $L$ is now exposed to its baths, the dissipative dynamics due to $\mathcal{L}_L(\cdot)$ forces the qutrit to restore its equilibrium state, $p_1^\prime \to p_1$ and $p_2^\prime \to p_2$. This, in turn, increases the energy of $L$ by absorbing some heat from the hot and cold baths or, equivalently, from the synthetic bath. Similarly, if $R$ is exposed to its baths, some of its energy is released to its synthetic bath in the form of heat and thereby attains its equilibrium. Note in this process, to reach the equilibrium, the populations of $\ket{3}$ do not remain constant throughout in both $L$ and $R$.

From the above arguments, we see that the unitary evolution with the interaction between $L$ and $R$ drives the composite out of equilibrium leading to a spontaneous heat flow from $L$ to $R$ for $-\beta_{LS}>-\beta_{RS}$. While, the dissipative evolution due to thermal interactions with the baths tries to restore the composite back to the initial equilibrium state ($\rho_L \otimes \rho_R$) by pumping some heat into $L$ and absorbing some heat from $R$. When both unitary and dissipative evolutions occur simultaneously, as in Eq.~\eqref{eq:2QutritME}, the opposing tendencies balance each other and result in a steady state, say $\sigma_{LR}$. This steady state is again block-diagonal in total energy eigenstates and has off-diagonal elements in the eigenstates $\ket{12}$ and $\ket{21}$. Nevertheless, the steady-state dynamics generate a heat flux from $L$ to $R$,
\begin{align}
 \dot{Q}_L=\Tr[\mathcal{L}_L (\sigma_{LR}) \ H_L]=\Tr[\mathcal{L}_L (\sigma_{L}) \ H_L] >0,  \end{align}
and $\dot{Q}_R=-\dot{Q}_L$. The expression of $\dot{Q}_L$ can be given analytically, and it has complicated dependencies with all the parameters. Rather, a numerical analysis is more illuminating; we have done so in Fig.~\ref{fig:KP}(a). Again for $-\beta_{LS}<-\beta_{RS}$, we find that $\dot{Q}_L <0$. This means that there is a spontaneous heat flow from $R$ to $L$.

In thermodynamics, the Kelvin-Planck statement of the second law states that heat can only flow from a hot bath to a cold bath when no external work is performed. As we see above for $-\beta_{LS}>-\beta_{RS}$, there is a heat flow from $L$ to $R$. This means that: (1) baths with negative inverse temperatures are 'hotter' than the baths with positive inverse temperatures; (2) baths with larger negative inverse temperatures are 'hotter' than those with smaller negative inverse temperatures. This is in conformation with the findings of Ramsey \cite{Ramsey1956}. However, one important point to be noted here is that although there is a heat flow from a negative to a positive temperature bath, the entropy flow is the opposite. This is indeed a signature of a bath having a negative temperature.
 
The Clausius statement, another formulation of the second law, states that the overall entropy production is always positive in a thermodynamical process. For steady-state thermodynamics with synthetic baths, the overall entropy production rate is given by
\begin{align}
    \Sigma = \dot{S}_{LR} -\beta_{LS}\dot{Q}_{L} - \beta_{RS}\dot{Q}_{R},
\end{align}
where $\dot{S}_{LR}=\partial S_{LR}/\partial t$ is the rate of change in von Neumann entropy $S_{LR}=-\Tr[\sigma_{LR} \log \sigma_{LR}]$, and $\beta_{LS}$ ($\beta_{RS}$) is the synthetic inverse temperature of $L$ ($R$) and $\dot{Q}_{L}$ ($\dot{Q}_{R}$) is the heat flux from $L$ ($R$). Note at steady state, the $\partial S_{LR}/\partial t=0$ and $\Sigma=(\beta_{RS}-\beta_{LS}) \ \dot{Q}_{L}$, as $\dot{Q}_{L}=-\dot{Q}_{R}$. For $-\beta_{LS}>-\beta_{RS}$, the heat flux from $L$ is positive, $\dot{Q}_L > 0$. Consequently, $\Sigma > 0$. Similarly, $\dot{Q}_L < 0$ for $-\beta_{LS}<-\beta_{RS}$, and thus $\Sigma > 0$. For $\beta_{LS}=\beta_{RS}$, we have $\Sigma=0$. Thus, the Clausius inequality in the differential form is
\begin{align*}
    \Sigma \geq 0,
\end{align*}
and it is always satisfied as long as the Kelvin-Planck statement is respected.

At steady state, the entropy production rate is positive, $\Sigma \geq 0$. This is mainly due to the dissipative interaction between the baths and the system. However, one may find out an entropy flow through the system $LR$, as
\begin{align*}
 \dot{S}_{X}=-\Tr[\mathcal{L}_X(\sigma_{LR}) \log \sigma_{LR}],   
\end{align*}
for $X=L,R$ (see Appendix~\ref{AppSec:SteadyState}). At steady state, $\dot{S}_{L}+\dot{S}_{R}=0$ as the state does not evolve over time. $\dot{S}_{L}>0$ implies that there is an entropy flux from bath with inverse temperature $\beta_{LS}$ to $R$ via $L$, and similarly for $\dot{S}_{R}>0$. In general, for a bath with positive temperature, an outflow of heat is associated to a decrease in entropy. One striking feature we must note here is that, although there is a spontaneous heat flow from a bath with negative temperature to a bath with positive temperature, the entropy flow is just opposite to that (see Fig.~\ref{fig:KP}(b)). This is also true when both baths are of negative temperatures. For a bath with negative temperature, an outflow of heat is associated with an increase in entropy of the bath. Thus, a bath with negative temperature in general acts as an entropy sink.    

\section{Quantum heat engines with a bath at negative temperature \label{sec:QHE}}
\begin{figure}
\includegraphics[width=0.95\columnwidth]{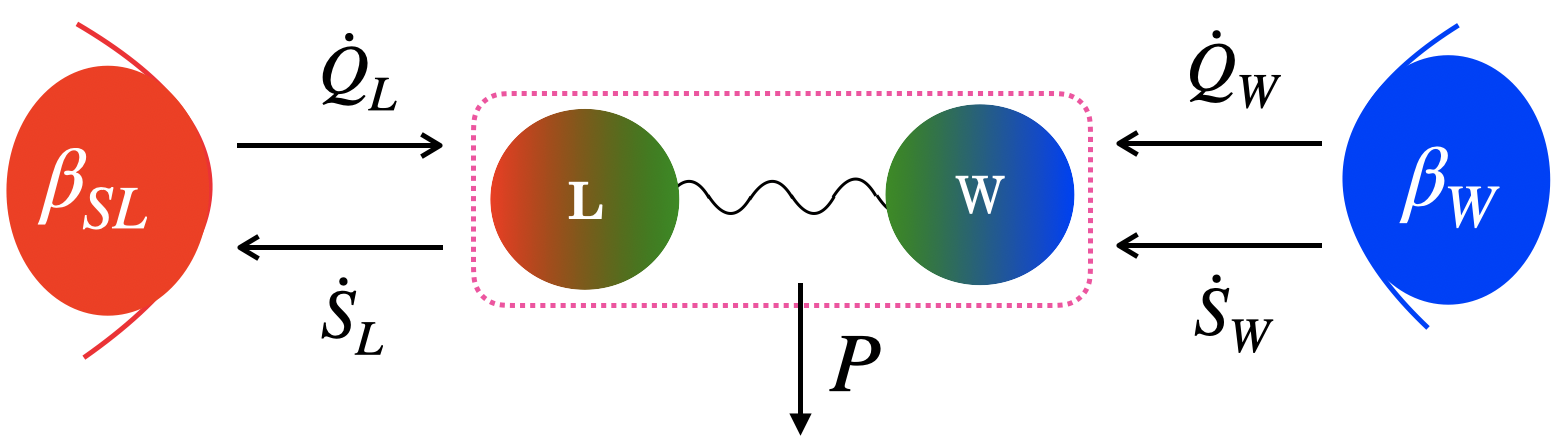}
\caption{A schematic of a quantum heat engine operating with two baths; one with negative inverse temperature $\beta_{SL}$ and the other with positive temperature $\beta_W$. The negative temperature is synthesized by weakly coupling a qutrit $L$ with a hot ($H$) and a cold ($C$) bath, as discussed in Fig.~\ref{fig:SynthBath}. This is as if the energy levels $\ket{1}$ and $\ket{2}$ are coupled to a synthetic bath at inverse temperature $\beta_{SL}$.  A qubit $W$ is weakly coupled to the bath with $\beta_W$. In the engine, $L$ and $W$ are coupled through a time-dependent interaction as in Eq.~\eqref{eq:Hdrive}. The arrows represent the direction of heat ($\dot{Q}_X$) and entropy ($\dot{S}_X$) fluxes for $X=L, W$, and $P$ represents the power of the engine. See text for more details. \label{fig:Devices} }
\end{figure}

Now we discuss heat engines operating with a bath at synthetic temperatures, particularly at negative temperatures. A device acting as a heat engine aims to transform heat into work. A generic heat engine consists of three primary parts: two separate heat baths with different temperatures and a working system. It operates by absorbing heat from the hot bath. The working system transforms part of this heat into work, dumping the rest into the cold bath. The model of a quantum heat engine (QHE) we are concerned with utilizes a synthetic bath with negative temperature and a heat bath with positive temperature (as depicted in Fig.~\ref{fig:Devices}). The working system is composed of a qutrit ($L$) and a qubit ($W$). The synthetic bath with inverse temperature $\beta_{LS}$ is created using two baths at different temperatures $\beta_H$ and $\beta_C$ and letting these weakly interact with $L$, similar to the one considered in Fig.~\ref{fig:SynthBath}. The Hamiltonian of $L$ is given by $H_L=(E_H-E_C)\proj{2} + E_H \proj{3}$. The qubit $W$ is weakly coupled to a bath at inverse temperature $\beta_W \geq 0$, and its Hamiltonian is $H_W=E_W \proj{2}$, where $E_W=E_H-E_C$. To operate the device as a heat engine, a time-dependent interaction is introduced between $L$ and $W$ driven by the interaction Hamiltonian
\begin{align}
 H_{in}^E (t)=\delta \ (\ket{11}\bra{22} \  e^{i \omega t } + \ket{22}\bra{11} \ e^{-i\omega t }). \label{eq:Hdrive} 
\end{align}
The total Hamiltonian is then $H_T^E=H_0 + H_{in}^E (t)$, where $H_0=H_L + H_W$. The overall dynamics will never lead to a steady state for a time-dependent interaction. However, for a periodic time dependence, there is a rotating frame in which the interaction becomes time-independent. For instance, to move from the laboratory frame to a rotating frame to we may introduce a rotation, given by the unitary $U=\exp[i H_{R} t]$ which satisfies $[H_{R}, H_0]=0$. For a suitable $H_{R}$, the interaction becomes time-independent, i.e., $V_{in}=U H_{in}^E (t) U^\dag$. In this rotating frame, the overall dynamics comprising the unitary and the dissipative evolution is given by (see Appendix~\ref{AppSec:SteadyState})
\begin{align}
 \dot{\rho}_{LW}^{R}=i [\rho_{LW}^{R}, \bar{H}_T^{E}] + \mathcal{L}_{L}(\rho_{LW}^{R}) + \mathcal{L}_{W}(\rho_{LW}^{R}),    
\end{align}
for a state $\rho_{LW}$, with ${\rho}_{LW}^{R}=U\rho_{LW}U^\dag$ and $\bar{H}_T^{E}=H_0 - H_{R}+V_{in}$. Note the LSOs remain unchanged in the rotating frame. Now that the time-dependence in the Hamiltonian is lifted, the dynamics attains a steady state $\sigma_{LW}^{R}$ in the rotating frame.

With this, the heat flux, entropy flux, and the power in the laboratory frame are quantified as (see Appendix~\ref{AppSec:SteadyState})
\begin{align}
 \dot{Q}_{X}=\tr[\mathcal{L}_X & (\sigma_{LW}^{R}) \ H_0], \ \ \dot{S}_X= - \Tr [\mathcal{L}_X(\rho_{LW}^{R})  \log \sigma_{LW}^R ], \nonumber \\
 \mbox{and} \ \ & P=i \ \tr[\sigma_{LW}^{R} \ [V_{in}, H_0]]. \end{align}
for $X=L,W$, and $[A,B]=AB-BA$. Here $\dot{Q}_{X}$ and $\dot{S}_{X}$ represent the heat and entropy fluxes through system $X$ respectively, and $P$ represents the power. The condition $\dot{Q}_{L}+\dot{Q}_{W}+P=0$ is always satisfied as required by the first law \cite{Alicki1979, Kosloff2014} and at steady state, $\dot{S}_{L}+\dot{S}_{W}=0$. For any negative inverse temperature $\beta_{LS}<0$, we have, as confirmed by numerical analysis, $\dot{Q}_{L}>0$, $\dot{Q}_{W}>0$, and $P<0$. This means the device draws heat from both the synthetic bath and the bath with inverse temperature $\beta_W$. For traditional engines, the efficiency is calculated as the ratio of work extracted and the heat absorbed by the engine from the hot bath. Here, the heat is absorbed from both baths. For each bath, the corresponding engine efficiency may be defined as
 \begin{align*}
  \eta_L= \frac{-P}{\dot{Q}_{L}}, \ \ \mbox{and} \ \  \eta_W= \frac{-P}{\dot{Q}_{W}}.
 \end{align*}
It can be easily checked that $\eta_L >1$ and $\eta_W >1$. Thus, the efficiency exceeds unity for an engine operating between baths with positive and negative temperatures. This is what is also claimed in literature \cite{Geusic1967}. We, however, find this conclusion incomplete. The heat-to-work conversion efficiency should always be defined with respect to the total amount of heat entering the engine and the amount of work produced out of that. In that sense, the total heat flux entering the engine is $\dot{Q}_{L}+\dot{Q}_{W}$, and this entire heat is converted into work. As a result, we find
 \begin{align*}
     \eta=\frac{-P}{\dot{Q}_{L}+\dot{Q}_{W}}=1,
 \end{align*}
i.e., the engine efficiency becomes unity. Thus, the efficiency of an engine can never exceed unity in any circumstance as long as the first law, i.e., the overall energy conservation, is respected. 

\begin{figure}
\includegraphics[width=0.95\columnwidth]{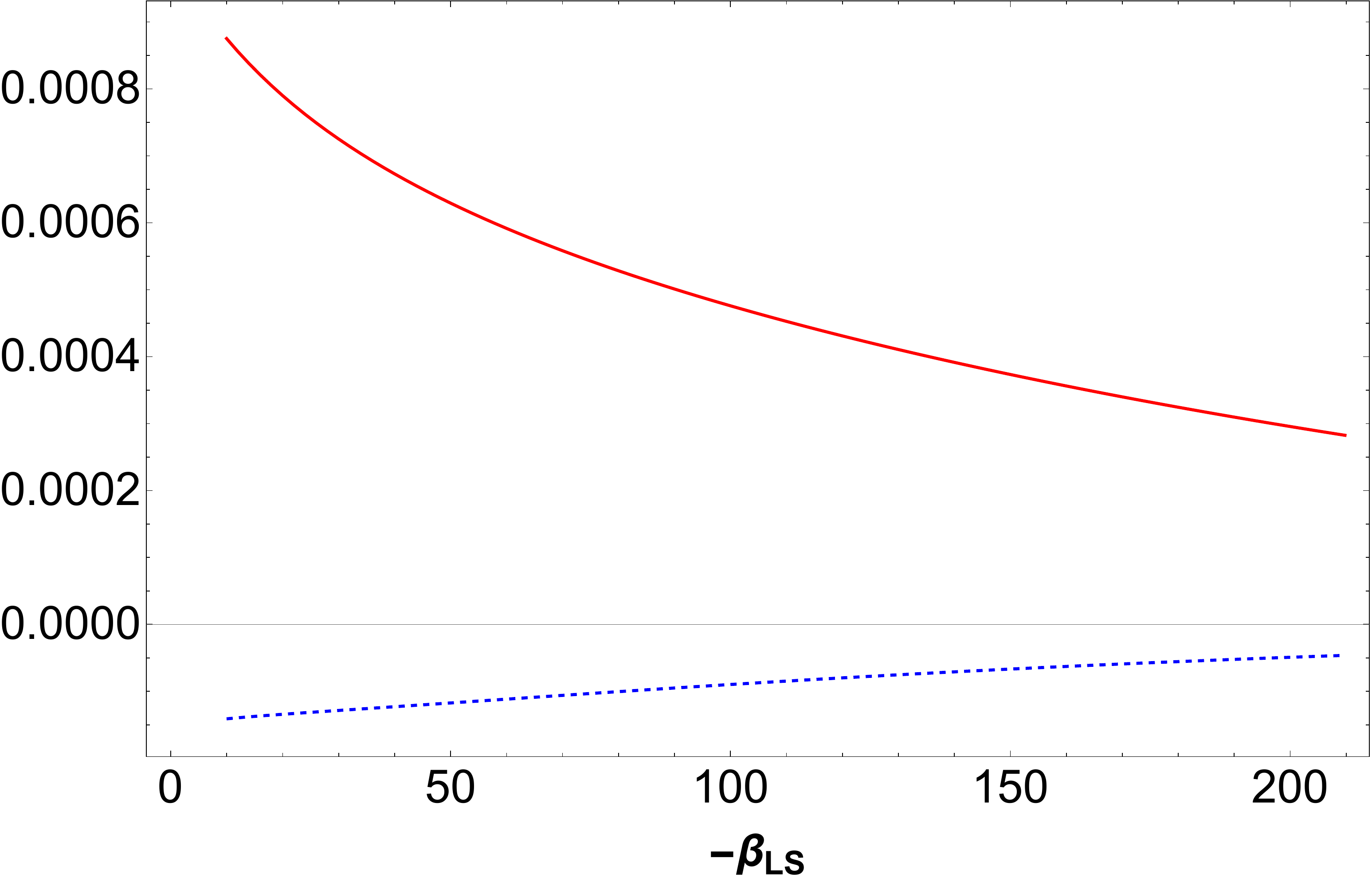}
\caption{Heat and entropy fluxes in the engine. The red-solid and blue-solid traces represent the change in heat flux ($\dot{Q}_L$) and entropy flux ($\dot{S}_L$) through $L$ in the engine with respect to $\beta_{SL}$ respectively. The calculation is done with the parameters: $\beta_H=0.01$, $\beta_W=0.1$, $\beta_C=10$, $\delta=5$, $\omega=1$, and $E_S=5$. As clearly seen, $\dot{Q}_L>0$ and $\dot{S}_L<0$ for all $\beta_{SL}<0$. See text for more details.} \label{fig:HeatEntropyFluxQHE} 
\end{figure}

This is, nevertheless, different from what we see in traditional heat engines. It raises a question on the physical meaning of the heat released or absorbed by a bath with a negative temperature. By definition, heat is a form of energy that is always associated with a change in the entropy of the corresponding bath. Heat flow, thus, occurs with an entropy flow. The work, on the other hand, is a pure form of energy and is not associated with any flow of entropy. In traditional engines operating with baths at positive temperatures, the direction of heat flow and the direction of entropy flow is the same. That is how a bath gets cooled down when it releases some heat. For a bath with a negative temperature, this is not true (see Fig.~\ref{fig:KP}(b)). There the bath's entropy increases as it releases heat. For the engine we have considered above, the direction of heat flux $\dot{Q}_{L}>0$ is opposite to the direction of entropy flux $\dot{S}_{L}<0$ in $L$ which is coupled to the negative bath (see Fig.~\ref{fig:HeatEntropyFluxQHE}). While for $W$, coupled to positive temperature bath, the direction of heat flux $\dot{Q}_{W}>0$ is the same as the direction of entropy flux $\dot{S}_{W}>0$. In fact, while heat is entering the engine from both baths, there is an entropy flow from the bath with inverse temperature $\beta_W$ to the bath with $\beta_{SL}$ where the latter acts as an entropy sink. 

In order to understand the thermodynamic nature of this heat, let us make a closer inspection of the process that is happening on $L$ and $W$ alone in the rotating frame. To create a synthetic bath with negative temperature, $L$ interacts with hot and cold baths with inverse temperatures $\beta_H$ and $\beta_C$. The corresponding heat fluxes are $\dot{Q}_{LH}=\Tr[\mathcal{L}_{LH} (\sigma_{LW}^R) \ H_L]$ and $\dot{Q}_{LC}=\Tr[\mathcal{L}_{LC} (\sigma_{LW}^R) \ H_L]$. Recall that $\mathcal{L}_{L}(\cdot)=\mathcal{L}_{LH}(\cdot)+\mathcal{L}_{LC}(\cdot)$. We can also quantify the power as $P_L=i \Tr[\sigma_{LW} \ [V_{in}, H_L]]$ which is produced in $L$. For the interaction $V_{in}$, we always find $\dot{Q}_{LH} >0$, $\dot{Q}_{LC}<0$, and $P_L>0$. For $W$, the heat flux and power can be similarly calculated, and they are $\dot{Q}_{W}>0$ and $P_W <0$. At steady state, we have 
\begin{align*}
& \dot{Q}_{LH}+\dot{Q}_{LC}+P_L=0, \\
& \dot{Q}_{W} + P_W=0, \\
& P=P_L+P_W.
\end{align*}
Note $\dot{Q}_{L}=\dot{Q}_{LH}+\dot{Q}_{LC}$ and $\dot{Q}_{L}=-P_L$. This may imply that the heat flux from the synthetic bath is quantitatively equal to the power extracted from $L$. But, as mentioned above, this cannot be just power, as power is not associated with any entropy flux. For justification, we may remove the interaction between qutrit $L$ and qubit $W$ and drive $L$ with an oscillating external field instead. This is exactly what is considered in \cite{Boukobza2007}, where power is quantified as the flux of pure energy being stored in the external field and is not associated with entropy flow (see Appendix~\ref{AppSec:SSDEng}). Here, on the contrary, we see an entropy flux opposite to the power extracted (heat flux) in $L$. This leads us to conclude that heat from a bath with a negative temperature is thermodynamic work but with negative entropy.

\section{Conclusion }
In this article, we have studied steady-state quantum thermodynamics with negative temperatures. For that, we have engineered synthetic baths by utilizing two baths at different positive temperatures and letting them weakly interact with qutrit systems in a particular fashion. These synthetic baths can assume arbitrary temperatures, including negative ones. These baths with negative temperatures are exploited to study steady-state thermodynamics. We have explored the thermodynamic laws, particularly the zeroth and second laws. We have shown that whenever two synthetic baths with identical temperatures are brought in contact, there is no heat flow in between. This, in turn, legitimizes the notion of temperatures in synthetic baths. On the other hand, for non-identical temperatures, there is a heat flow. Further, heat always flows spontaneously from a bath with a negative temperature to a positive one. This implies that a bath with a negative temperature is `hotter' than a bath with a positive temperature. Further, there is a heat flow from a bath with a less negative temperature to a bath with a more negative temperature. We have then studied the Clausius statement of the second law in case of negative temperatures and found that there is always non-negative entropy production in the steady-state thermal processes. Then, we have introduced a heat engine model that operates between two baths, one with negative temperature and the other with positive temperature. Unlike traditional engines, these engines always yield unit heat-to-work conversion efficiency. A systematic analysis has revealed that the heat flow from a bath with a negative temperature is equivalent to an injection of work into the working system by an equal amount. This is exactly the reason why these engines yield unit efficiency.  

The proposed engines utilizes qutrit and qubit systems and two or three baths at different temperatures. Thus, these engines can be realized experimentally, for example, using atom-optical \cite{Johannes2016, Klatzow2019, Wang2022} and NMR \cite{Peterson2019} systems. Finally, we conclude that:

\begin{itemize}
\item A thermal bath with a negative temperature can be synthesized with two baths with different positive temperatures.

\item In steady-state thermodynamics with negative temperatures, the zeroth law and the Clausius statement of the second law remain unchanged. However, the Kelvin-Plack statement is to be appended to incorporate that there is a spontaneous heat flow from a bath with a negative temperature to a bath with a positive temperature and from a bath with smaller negative temperature to a bath with a larger negative temperature.

\item A continuous heat engine can be constructed using baths with negative and positive temperatures. In such engines, the heat-to-work conversion efficiency is always unity. This is maximum for any device that respects first law, i.e., conservation of total energy.

\item On the fundamental level, the thermodynamic nature of heat from a bath with a positive temperature is qualitatively different from the one with a negative temperature. For the former, heat flows in the same direction as entropy flow. For the latter, heat flows in the opposite direction of entropy flow. Further, heat from a bath with a negative temperature is thermodynamic work but with negative entropy.  
\end{itemize}

\begin{acknowledgments}
M.L.B., U.B., and M.L. thankfully acknowledge support from: ERC AdG NOQIA; Ministerio de Ciencia y Innovation Agencia Estatal de Investigaciones (PGC2018-097027-B-I00/10.13039/501100011033, CEX2019-000910-S/10.13039/501100011033, Plan National FIDEUA PID2019-106901GB-I00, FPI, QUANTERA MAQS PCI2019-111828-2, QUANTERA DYNAMITE PCI2022-132919, Proyectos de I+D+I ``Retos Colaboraci\'on" QUSPIN RTC2019-007196-7); MICIIN with funding from European Union NextGenerationEU(PRTR-C17.I1) and by Generalitat de Catalunya; Fundaci\'o Cellex; Fundaci\'o Mir-Puig; Generalitat de Catalunya (European Social Fund FEDER and CERCA program, AGAUR Grant No. 2021 SGR 01452, QuantumCAT \ U16-011424, co-funded by ERDF Operational Program of Catalonia 2014-2020); Barcelona Supercomputing Center MareNostrum (FI-2022-1-0042); EU (PASQuanS2.1, 101113690); EU Horizon 2020 FET-OPEN OPTOlogic (Grant No 899794); EU Horizon Europe Program (Grant Agreement 101080086 - NeQST), National Science Centre, Poland (Symfonia Grant No. 2016/20/W/ST4/00314); ICFO Internal ``QuantumGaud" project; European Union's Horizon 2020 research and innovation program under the Marie-Sklodowska-Curie grant agreement No 101029393 (STREDCH) and No 847648 (``La Caixa" Junior Leaders fellowships ID100010434:LCF/BQ/PI19/ 11690013,LCF/ BQ/PI20/ 11760031, LCF/BQ/PR20/11770012,LCF/BQ/ PR21/11840013). M.L.B acknowledges the financial support from MCIN/AEI/10.13039/5011000  11033. V.S. acknowledges the financial support from the Institute for Basic Science (IBS) in the Republic of Korea through the project IBS-R024-D1. M.N.B. gratefully acknowledges financial support from SERB-DST (CRG/2019/002199), Government of India.
\end{acknowledgments}

%\bibliography{neg_tem.bib}

%merlin.mbs apsrev4-1.bst 2010-07-25 4.21a (PWD, AO, DPC) hacked
%Control: key (0)
%Control: author (0) dotless jnrlst
%Control: editor formatted (1) identically to author
%Control: production of article title (0) allowed
%Control: page (1) range
%Control: year (0) verbatim
%Control: production of eprint (0) enabled
%

\

%\onecolumngrid
%\newpage
\appendix

\section{Rotating frame and steady-state thermodynamics \label{AppSec:SteadyState}}
Let us consider the setup discussed in Section~\ref{sec:QHE} of the main text. The Hamiltonian of the working systems are
\begin{align*}
 & H_L=(E_H-E_C)\proj{2} + E_H \proj{3}, \\ 
 & H_W=E_W \proj{2},\\ 
 & H_{in}^E (t)=\delta \ (\ket{11}\bra{22} \  e^{i \omega t } + \ket{22}\bra{11} \ e^{-i\omega t }), \\ 
 & H_T^E(t)=H_L + H_W + H_{in}^E (t),
\end{align*}
where $E_W=E_H-E_C$. The Hamiltonian $H_L$ corresponds to a qutrit $L$. It weakly interacts with a hot ($H$) and a cold ($C$) bath at inverse temperatures $\beta_H$ and $\beta_C$ (as described in Section~\ref{sec:QHE} of the main text) to synthesize a bath with negative inverse temperature $\beta_{LS}$. The qubit $W$, with Hamiltonian $H_W$, weakly interacts with a bath with positive inverse temperature $\beta_W$. The $L$ and $W$ interact between them with a time-dependent interaction Hamiltonian given by $H_{in}^E (t)$. After having all these interactions, the overall dynamics of the composite $LW$ is 
\begin{align}
 \frac{\partial \rho_{LW}(t)}{\partial t}=i [\rho_{LW}(t), H_T^{E}(t)] + \mathcal{L}_{L}(\rho_{LW}(t)) + \mathcal{L}_{W}(\rho_{LW}(t)),    
\end{align}
for a state $\rho_{LW}$, where $\mathcal{L}_{L}(\cdot)=\mathcal{L}_{LH}(\cdot)+\mathcal{L}_{LC}(\cdot)$ is the Lindbld super-operator (LOS) representing the dissipative dynamics due to baths $H$ and $C$, and $\mathcal{L}_{W}(\cdot)$ is the LOS for the bath with inverse temperature $\beta_W$. For this dynamics, the heat flux and power are defined as \cite{Alicki1979, Boukobza2006a}
\begin{align}
    \dot{Q}&=\Tr\left[\frac{\partial \rho_{LW}(t)}{\partial t} \ H_T^E(t)\right], \\ 
    P&= \Tr \left[\rho_{LW} \ \frac{\partial H_T^E(t)}{\partial t} \right].
\end{align}
Note the heat flux $\dot{Q}$ and the power $P$ may have time dependence. 

For time-dependent Hamiltonians, the dynamics generally never leads to a steady state. However, for a periodic time-dependence, as in $H_{in}^E (t)$, there is a rotating frame in which the Hamiltonian can be made time-independent. For that, a counter-rotation is applied on the laboratory frame by $U=e^{i H_R t}$ with $[H_R, H_0]$, where $H_0=H_L+H_W$. In the rotating frame, an operator $A$ in the laboratory frame transforms as $A \to (A)_R=U A U^\dag$. Further, there exists a Hamiltonian $H_R$ for which the interaction Hamiltonian reduces to a time-independent one, given by $V_{in}=U H_{in}^E (t) U^\dag$. Accordingly, the overall Hamiltonian becomes time-independent, and it is $\bar{H}^E_T=H_0-H_R+V_{in}$. In this rotating frame, the overall dynamics is recast as
\begin{align}
 \frac{\partial \rho_{LW}^R(t)}{\partial t}= \mathcal{L}_U(\rho_{LW}^{R}) + \mathcal{L}_{L}(\rho_{LW}^{R}) + \mathcal{L}_{W}(\rho_{LW}^{R}), \label{AppEq:RotDyn}    
\end{align}
for a state $\rho_{LW}$, with ${\rho}_{LW}^{R}=U\rho_{LW}U^\dag$. Here we denote $\mathcal{L}_U(\rho_{LW}^{R})=i [\rho_{LW}^{R}, \bar{H}_T^{E}]$ which is the unitary controbution to the dynamics. This dynamics can lead to a steady state, say $\sigma^R_{LW}$. It can be easily checked that the LOSs remain unchanged in this rotating frame. Given that $\Tr[AB]=\Tr[(A)_R(B)_R]$ for two arbitrary operators $A$ and $B$, we may re-express the heat flux and power as \cite{Boukobza2006a}
\begin{align}
\dot{Q}=\Tr\left[ \left(\frac{\partial \rho_{LW}(t)}{\partial t}\right)_R \ \left(H_T^E(t)\right)_R\right] = \Tr\left[\mathcal{L}(\rho_{LW}^{R}) \ H_0 \right], 
\end{align}
where $\mathcal{L}(\rho_{LW}^{R})=\mathcal{L}_{L}(\rho_{LW}^{R}) + \mathcal{L}_{W}(\rho_{LW}^{R})$, and the power as 
\begin{align}
    P= \Tr \left[\left(\rho_{LW}\right)_R \ \left(\frac{\partial H_T^E(t)}{\partial t}\right)_R \right]= - i \Tr \left[\rho_{LW}^{R} \ [H_0, V_{in}] \right], 
\end{align}
where $[A,B]=AB-BA$. The heat flux $\dot{Q}$ can be divided into two parts. One contribution comes from interaction of $L$ with baths $H$ and $C$, i.e., $\dot{Q}_L=\Tr\left[\mathcal{L}_L(\rho_{LW}^{R}) \ H_0 \right]$ and the other due to interaction between $W$ with its bath, i.e., $\dot{Q}_W=\Tr\left[\mathcal{L}_W(\rho_{LW}^{R}) \ H_0 \right]$. 

Now we study the entropy flux through $LW$. Note the rate of change in von Neumann entropy is given by
\begin{align*}
 \dot{S} & =- \Tr \left[\frac{\partial \rho_{LW}(t)}{\partial t} \log \rho_{LW}(t)  \right]  \\ 
 &= - \Tr \left[ \left(\frac{\partial \rho_{LW}(t)}{\partial t}\right)_R \log \left(\rho_{LW}(t) \right)_R \right].   
\end{align*}
In the rotating frame and at steady-state, it reduces to 
\begin{align*}
 \dot{S}= - \Tr \left[ \frac{\partial \sigma_{LW}^R}{\partial t} \log \sigma_{LW}^R \right] =0.  
\end{align*}
The rate of change in entropy vanishes because the state does not change over time. Nevertheless, there still can be a non-vanishing flux of entropy passing through $L$ and $W$. Given that unitary dynamics does not contribute to the entropy flux, i.e.,  $- \Tr [\mathcal{L}_U(\rho_{LW}^{R})  \log \sigma_{LW}^R ] =0$, we can calculate the entropy flux through $L$ and $W$ respectively as
\begin{align}
& \dot{S}_L= - \Tr [\mathcal{L}_L(\rho_{LW}^{R})  \log \sigma_{LW}^R ], \\
& \dot{S}_W= - \Tr [\mathcal{L}_W(\rho_{LW}^{R})  \log \sigma_{LW}^R ],
\end{align}
where $\dot{S}_L + \dot{S}_R=0$. 

\section{Power and associated entropy flux in traditional steady-state engines \label{AppSec:SSDEng}}
In traditional engines, unlike heat, work is a form of energy not associated with entropy. To understand that, we may reconsider the heat engine composed of a qutrit and two baths at different temperatures, which is also studied in \cite{Boukobza2007}. Consider the qutrit system outlined in Section~\ref{Sec:SynthBath} where the Hamiltonian of the qutrit is $H_0=(E_H-E_C)\proj{2} + E_H \proj{3}$, and it is weakly interacting with a hot ($H$) and cold ($C$) with inverse temperatures $\beta_H$ and $\beta_C$ respectively. In addition, an external driving 
\begin{align}
 H_d(t)= \alpha \ (\ket{1}\bra{2} e^{i \omega t} + \ket{2}\bra{1} e^{-i \omega t})   
\end{align}
is introduced. As a result, the reduced dynamics becomes
\begin{align}
    \dot{\rho}= i \  [\rho, H_T(t)] + \mathcal{L}_H (\rho) + \mathcal{L}_H (\rho),
\end{align}
where $H_T(t)=H_0+H_d(t)$, and $\mathcal{L}_H (\cdot)$ and $\mathcal{L}_C (\cdot)$ are LSOs representing the dissipation due to interactions with hot and cold baths. Again, as discussed in Appendix~\ref{AppSec:SteadyState}, this dynamics leads to a steady-state, say $\sigma^R$, in a rotating frame. The heat fluxes, entropy fluxes, and power are calculated respectively as
\begin{align*}
 \dot{Q}_H&=\tr \left[\mathcal{L}_H(\sigma^R) H_0 \right], \\
 \dot{Q}_C&=\tr \left[\mathcal{L}_C(\sigma^R) H_0 \right], \\
 \dot{S}_H&=-\tr \left[\mathcal{L}_H(\sigma^R) \log \sigma^R \right], \\
 \dot{S}_C&=-\tr \left[\mathcal{L}_C(\sigma^R) \log \sigma^R \right], \\
 P&= - i \tr \left[\sigma^R \ [H_0, V] \right],
\end{align*}
where $V=\alpha \ (\ket{1}\bra{2} + \ket{2}\bra{1})$. Here $\dot{Q}_X$ and $\dot{S}_X$ represent the heat and entropy fluxes from bath $X=H, C$, and $P$ is the power. As required by first law, $\dot{Q}_H+\dot{Q}_C+P=0$ and, at steady-state, $\dot{S}_H=-\dot{S}_C$. When the device operates as an engine, we have $\dot{Q}_H>0$, $\dot{Q}_C<0$, and $P<0$. In this case, heat enters from the hot bath. Part of that heat is converted into work which is stored in the driving field, and the rest is dumped into the cold bath. Further, in this case, $\dot{S}_H>0$ and $\dot{S}_C<0$. Clearly, at steady-state, whatever amount of entropy enters the system from the hot bath is released to the cold bath. Thus the work-flux, i.e., power $P$, does not carry any entropy.

\end{document}